%% file: skct_accepted.tex
\documentclass[11pt,preprint]{aastex62}
\usepackage[utf8]{inputenc}
\usepackage{ifpdf}
\usepackage{hyperref}
\usepackage{amsmath}
\usepackage{natbib}

\newcommand{\ifms}[1]{}	
\newcommand{\ifpp}[1]{#1}	
\input units.tex

\begin{document}
\title{Improved Analysis of Clarke Exobelt Detectability}
\author[0000-0003-0251-1237]{Shauna Sallmen}
\affiliation{University of Wisconsin - La Crosse, La Crosse, WI 54601}
\author[0000-0001-8078-9395]{Eric J. Korpela}
\affiliation{University of California, Berkeley, CA 94720-7450}
\author{Kaisa Crawford-Taylor}
\affiliation{University of Wisconsin - La Crosse, La Crosse, WI 54601}

\begin{abstract}
We analyze the potential transit light curve effects due to a Clarke belt of satellites around an exoplanet. Building on code and analysis from \citet{ksg2015}, we refine the transit analysis of \citet{exobelt} by incorporating limb-darkening and taking an observer-centered approach to examining residuals.  These considerations make Clarke exobelt detectability more difficult than previous estimates.
We also consider practical dynamical issues for exobelts,  confirming that synchronously orbiting belts are dynamically unstable around planets in the habitable zones of M stars, and determining the maximum quasi-stable belt size in these situations.
Using simulations for both G and M stars, we conclude that to have an even marginally detectable impact on transit light curves, exobelts must be substantially denser than previous estimates. We also estimate collision rates for the required satellite densities assuming random orbits, and find they would present significant monitoring and guidance challenges.  If detectable belts exist, they would require some (possibly high) degree of ordering to avoid collisions, and must be actively maintained or they will dissipate on relatively short astronomical timescales.
We conclude that detectable exobelts are likely to be rare, and have extremely low prospects for detection by transit monitoring from both current and upcoming missions.
\end{abstract}

\section{Introduction}

The potential for detection of extraterrestrial intelligence through the effects of macroscopic artifacts transiting a star has been considered for more than a decade.  \citet{arnold05} showed that large geometric or periodic structures could generate detectable changes in transit profiles.  More recently \citet[][hereafter \KSG]{ksg2015}, showed that smaller structures, such as a cloud of mirrors or solar panels orbiting a planet, could be detected with \JWST\ in a reasonable number of transits.  \citet[][hereafter \SN]{exobelt} intriguingly extended that concept to Clarke exobelts, a band of satellites in synchronous orbit about a planet. \SN\ extrapolated the current increase in satellites in geosynchronous orbit around Earth into the future and estimated the impact of such an exobelt on planetary transit light curves. Although a valuable first step, there are several issues with that analysis of such a belt's detectability.

First, there are no stable synchronous orbits around tidally locked planets apart from the semi-stable Lagrange points \citep{lagrange1772}, L1 and L2, thus planets in the habitable zones (HZs) of K and M stars cannot have synchronous Clarke exobelts. Earth-like planets around F or G stars could still possess synchronous exobelts, so it is worth considering their detectability. Exobelts comprised of satellites that aren't in synchronous orbit may exist around an M star's habitable planets, although their smaller size will reduce their predicted impact on transit light curves.

The light-curve analysis in \SN\ is a good starting point, but contains a number of simplifications, some of which are critical to the potential detectability of such a system. First, the observability of an effect should be analyzed with the same methods used in observations. \SN\ assumes the planet-only transit light curve is a known quantity, and determines the effects of an exobelt. A more appropriate comparison is an isolated planet producing a transit of similar depth as the simulated one (vs. comparison to the same planet transiting without a belt). An even more observer-based approach is to fit a planet-only model to the simulated transit profile, and determine the remaining residuals. Potentially, even more of the exobelt's effects may be incorporated into this ``best-fit" model, further reducing the residuals.  We used both approaches in \KSG.

\SN\ also neglected the effects of limb-darkening.  When the star's fainter limb is considered, the exobelt's effects on transit ingress and egress will be significantly reduced.

In this paper we address these and other simplifications to create revised transit simulations and analyze their prospects for detection. In Section \ref{sect:stability} we look at the stability of orbits around tidally locked planets, in order to help guide some of the transit simulations described in \ref{sect:methsim}.  Section \ref{sect:resultsdiscussion} discusses the simulation results and implications of our analysis. 

\section{Methods}

\subsection{Orbit Stability}\label{sect:stability}
As noted by Socas-Navarro\footnote{https://www.hou.usra.edu/meetings/technosignatures2018/presentation/socasnavarro.mp4, retrieved 01 October 2019}, synchronous orbit Clarke exobelts are unstable if the planet is itself in synchronous rotation about its star. Specifically, the synchronous orbit location $r_{synch}$, calculated using only the planetary mass, is beyond the L1 Lagrange point $r_{L1}$, 
\begin{equation}
    r_{synch} = \frac{G}{4\pi^2}MT_{orb}^2 = 3^{1/3} r_{L1} 
\end{equation}
with $T_{orb} = T_{rotplanet}$. 
The discussion in \KSG\ reveals that habitable planets around G-type stars are not generally tidally locked into synchronous rotation, unless they are at the inner edge of the habitable zone. However, K and M stars with advanced civilizations are most probably old enough ($\gtsim 1$ Gyr) that any habitable planets will have reached synchronous rotation (see references in \KSG). Thus synchronous Clarke exobelts are extremely unlikely to be found around such planets. 

To evaluate the maximum distance at which a satellite could orbit relatively stably around a planet tidally locked into synchronous rotation, we used the REBOUND N-body simulator \citep{ReinLiu2012} with the adaptive, high-order integrator IAS15 \citep{ReinSpiegel2015}. We included the star, planet, and a satellite similar to \JWST\ (mass $\sim 6000$ kg, cross-sectional area $\sim 30$m$^2$). For such a small cross-sectional area and large mass, the effects of radiation and solar wind pressures will be small compared with the gravitational effects, so were not included. The circular orbit speed around an isolated planet defined the satellite's initial velocity.

We ran orbital simulations for an Earth-like ($R_p = R_{\oplus}$, $M_p = M_{\oplus}$) planet orbiting an M8 star ($M_{star} = 0.1M_{\odot}$) at the inner and outer edges of the habitable zone (0.023 AU and 0.063 AU, respectively), as determined using the optimistic values of \citet{kopparapuetal13,kopparapuetal14}, accessed via \href{http://depts.washington.edu/naivpl/content/hz-calculator}{http://depts.washington.edu/naivpl/content/hz-calculator}. If we assume the planet is tidally locked to always face its star, then at the inner edge of the habitable zone, a synchronously orbiting satellite would have $R = 16.77 R_p$. Our simulation confirms that a satellite starting at this distance quickly moves away from both the planet and the star, escaping the planet well before even one orbit would have elapsed.

\begin{figure}[tbp]
\ifpdf
\plotone{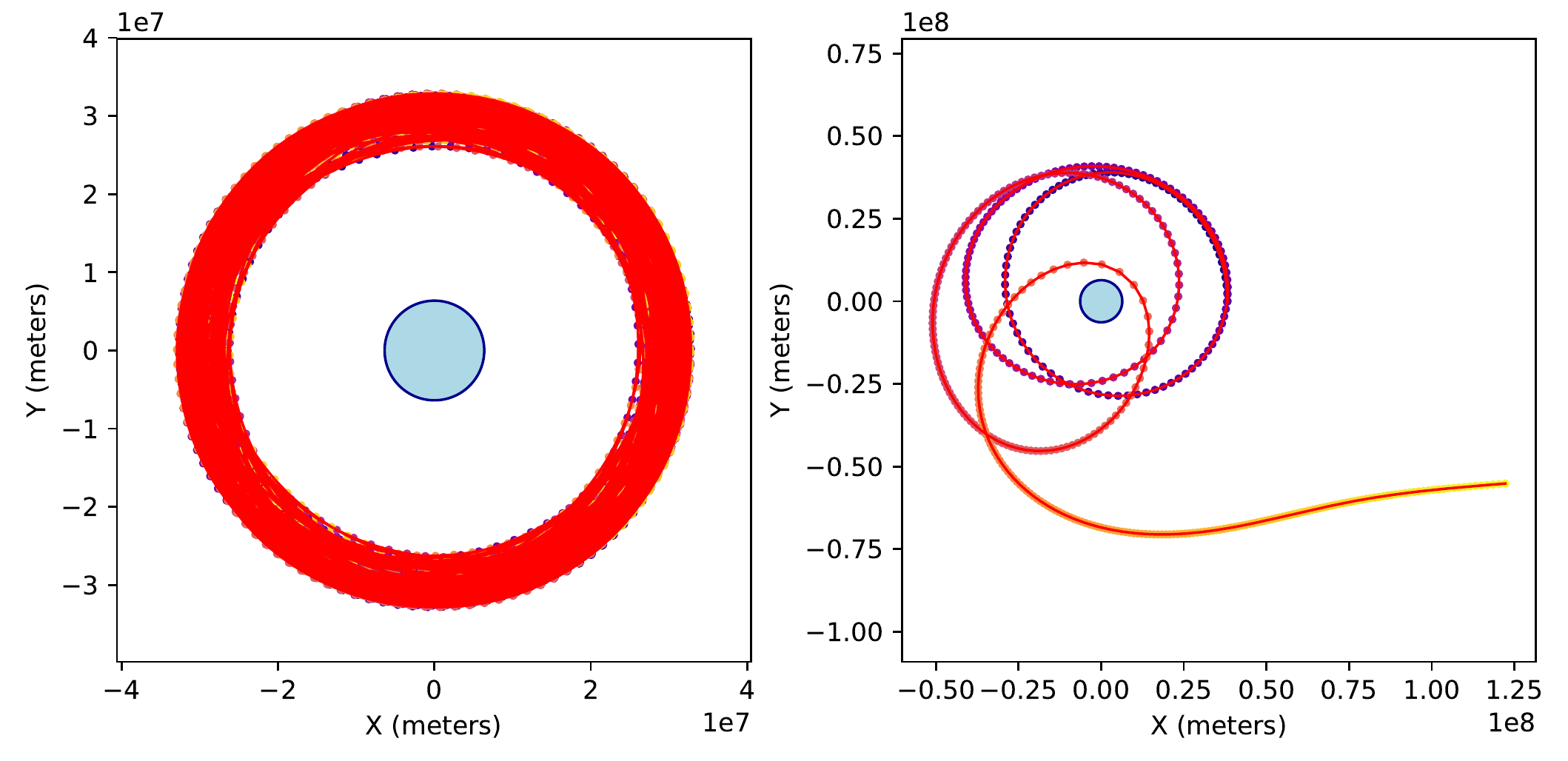}
\else
\plotone{M8_HZin_stability_XY.eps}
\fi
\caption{{\it Left}: Satellite orbiting an Earth analog that is at the inner edge of the HZ for an M8 star. Without the star's influence, the satellite would be in a circular orbit at 5 planetary (Earth) radii ($R_m = 5R_p$). The satellite orbits are shown in a reference frame centered on the planet (but not in the rotating reference frame). A red line connects the dots representing the satellite's position throughout the simulation. The dots' color gradually shifts from purple (start) to yellow (end) as time increments in the simulation. {\it Right}: Same, but with $R_m = 6R_p$.}
\label{fig:M8_HZin}
\end{figure}

The left panel of Figure \ref{fig:M8_HZin} shows 100 orbits of a satellite initially in a circular orbit 5 Earth radii in size.  This orbit is relatively stable and could be maintained with relatively minor adjustments. 
The right panel shows the situation for a satellite initially at 6 Earth radii. In this case, the satellite reaches escape velocity after fewer than 5 orbits, even though its orbit is less than half the size required for the synchronous orbit of a Clarke exobelt. 

For a synchronously rotating Earth-mass planet at the outer edge of the M8 habitable zone (0.063 AU), the synchronous satellite distance is $R_{\rm synch} = 45.9R_p$. However, an initially $14 R_p$ circular orbit is about as stable as that shown in Fig.~\ref{fig:M8_HZin}, but a satellite initially at $15 R_p$ escapes within 57 orbits.
Similarly, for an Earth analog in synchronous rotation orbiting an M5 star at the HZ's inner or outer edge, the last stable orbits are estimated to be at $R_{\rm sat} = 12R_p$ and $33R_p$, respectively. These are much smaller than the synchronous satellite orbit sizes of $42R_p$ and $110R_p$. For an M0 star, the most distant stable orbits are at 26 and 69 planet radii, compared with $87R_p$ and $225R_p$ for satellite orbits synchronous with both the planet's spin and orbit period. In all these cases, the maximum size relatively stable orbit is estimated to be $\sim 0.3R_{\rm synch}$.

Based on these orbital dynamics results, realistic exobelts must be significantly smaller in extent than those modeled by \SN.  These effects would also have significant consequences for rings or moons around tidally locked worlds.  Both must be tightly bound to the planet. We've used the estimates made above to guide our transit simulations in what follows. 

\subsection{Belt Transit Simulations}\label{sect:methsim}

\SN\ assumed a cylindrical surface to derive an analytic expression for belt absorptance. This is a reasonable first-order approximation, but a belt of satellites with very similar altitudes is actually a subsection of the thin spherical shell described in \KSG\ Section 4.3. To model the transit light curves of planets with exobelts, we adapted \KSG's brute force method (see Section 3) of simulating a satellite fleet distributed in a uniform-density thin spherical shell. This software uses IDL version 8.3 (Exelis Visual Information Solutions, Boulder, Colorado) to generate two-dimensional arrays for both the star's brightness and the planet+exobelt absorptance. The planet radius $R_p$ is always 100 pixels in the simulations. By shifting the relative positions of the star and absorptance arrays, we calculate the transmitted light during all phases of a transit. This 2-D pixel-based approach allows our models to include limb-darkening.  As noted earlier, including limb-darkening significantly changes the transit light curves.

For such a shell (inner radius $R_i$ to outer radius $R_o$), the absorptance is related to the projected density of the satellite fleet: 
\begin{equation}
\begin{split}
\rho_{\rm proj} = 2 \rho_{\rm sat} (\sqrt{R_o^2-r^2} - \sqrt{R_i^2-r^2}) \qquad (r \lt R_i) \\
\rho_{\rm proj} = 2 \rho_{\rm sat} (\sqrt{R_o^2-r^2}) \qquad (R_i \le r \le R_o) \\
\end{split}
\end{equation}
We assume satellites comprising the belts are at near-constant altitude, and set $R_o - R_i$ equal to one pixel of our simulation, with the belt at the desired distance $R_{belt}$ from the planet. To generate the belt absorptance, we restrict the thin shell absorptance profile to satellite orbit inclinations $< \pm \gamma$, to produce a belt height $2 R_{belt} \sin(\gamma)$. We also specify the absorptance $\chi_0$ of the near wall at the center of the belt. This matches the notation of \SN, where $\chi_0$ is the fraction of light blocked by a surface element when looking straight through one wall of the belt.

If a planet's spin and orbit axes are misaligned, then an equatorial satellite belt's transit effects depend on where in the orbit transit occurs. \SN\ analyzed the effects if transit occurs at a solstice, when a belt isn't viewed edge-on, so only part of the near and far belt walls overlap. At an equinox, the belt will appear edge-on without overlap. To model equinox transits, our code rotates the planet+belt absorptance array so that the edge-on belt approaches the star at an angle. This complements the solstice transit effects analyzed by \SN.

We extend \SN's analysis and model transiting planets whose orbits aren't precisely edge-on. We alter the impact parameter of our simulations, so that the planet+belt do not cross the center of the star. Stellar limb-darkening enhances the importance of this effect.

Note that an altitude spread of one pixel in our simulations ($\sim 64$km) is substantially thicker than Earth's belt, which \SN\ notes is $\sim 150$m thick. Both these values are small compared to the distance an orbiting planet would transit during an exposure and are unlikely to produce a relative difference. 

Stellar properties are determined from \citet{zombeck} with limb-darkening calculated using the interpolation of \citet{eastman13} over the values of \citet{ClaretBloemen11} (for values and details see Table 1 and Section 3.1 of \KSG.) 

This paper focuses on four situations.  Table \ref{table:SimProps} summarizes the properties of stars, planet orbits, and belt sizes used in each. Column 1 names the simulation. In the top half of the table, columns 2-6 contain the stellar properties as described above. In the bottom half, columns 3-4 contain the planet orbit size in AU and orbit period in days, while column 5 contains the distance of the belt in planetary radii.

\begin{table*}[tb]
\begin{center}
\caption{Simulation Parameters\label{table:SimProps}}
\begin{tabular}{ccccccccc}
\hline
Simulation & $T_{\rm eff}$ (K) & $M/M_{\odot}$ & $R/R_{\odot}$ & $L/L_{\odot}$
        & log($g$) \\
\hline\hline
G2HZin & 5780 & 1.00 & 1.00 & 1.00 & 4.35 \\
Earth-Sun & 5780 & 1.00 & 1.00 & 1.00 & 4.35 \\
M5HZin & 3120 & 0.21 & 0.32 & 0.0079 & 4.87 \\
M5HZout & 3120 & 0.21 & 0.32 & 0.0079 & 4.87\\
\hline\hline\ \\
Simulation & & $a_{\rm orb}$ (AU) & $P_{\rm orb}$ (d) & $R_{\rm belt}/R_{\oplus}$ & \\
\hline\hline
G2HZin & & 0.75 & 237.24 & 10 &  \\
Earth-Sun & & 1.00 & 365.25 & 6.6  &  \\
M5HZin & & 0.073 & 15.72 & 12 &   \\
M5HZout & & 0.19 & 66.01 & 33 &  \\
\hline
\end{tabular}
\end{center}
\end{table*}

For all of these,  the planet is an Earth analog with the same mass and size as Earth, and a maximum satellite orbit inclination of $\gamma = 15\degrees$ determines the belt height.
\begin{itemize}
    \item {\it G2HZin:} A belt of radius $R_{belt} = 10 R_p$ surrounding a planet at the inner edge of the habitable zone ($HZ_{in}$ = 0.75 AU) for a G2 star.
    \item {\it Earth-Sun:} Earth orbiting 1AU from a G2 solar abundance star with a geosynchronous belt at $R_{belt} = 6.6R_{\oplus}$.
    \item {\it M5HZin} and {\it M5HZout:} Planets orbiting an M5 star at the HZ's inner ($HZ_{in}$) and outer ($HZ_{out}$) edges for an M5 star, with satellites at the most distant location of a ``stable" satellite orbit for each.
\end{itemize}

\section{Results and Discussion}\label{sect:resultsdiscussion}

\subsection{Sun-like Stars}\label{sect:sunlike}
\subsubsection{Earth Analog at $HZ_{in}$ of G2 Star}\label{sect:G2HZin}
The first simulation (short-hand G2HZin) is directly comparable to a simulation from \KSG. That reference cloud of satellites surrounds an Earth analog at the inner edge of the HZ of a G2 star with half-solar abundance. For a constant-absorptance cloud of satellites extending to 10$R_p$ with an effective cross-sectional area (total absorptance) the same as the Earth analog, \KSG\ estimated that a minimum of 4 transits of \JWST\ observing time would be required for detection. This estimate was made by scaling the observed noise for Kepler-10 to account for \JWST's improved sensitivity, ignoring potential saturation issues. Kepler-10 is a relatively bright (Kp = 10.96, V$\sim$11.1) G2 star, at a distance of 173 $\pm$ 27 pc \citep{batalha2011}. \KSG\ used the anticipated peak transmission for the NIRCam F150W2 filter of 0.8, but the actual peak transmission of the filter is $\sim 0.45$\footnote{From  \href{http://svo2.cab.inta-csic.es/svo/theory/fps/}{http://svo2.cab.inta-csic.es/svo/theory/fps/}, retrieved 08 January 2019.}.  The \JWST\ Exposure Time  Calculator also shows that NIRCAM F150W2 will saturate on G stars brighter than V$\sim$15, and therefore \JWST\ observations of Kepler-10 will not be possible.  Unfortunately this means that the observation described herein, if performed on a V=15 G2 star (at $\sim 1000$ pc), would require 31 rather than 4 transits.  Such an observation seems unrealistic, given the 237-day orbital period for this reference case.

We have also looked into the capabilities of \TESS\ to handle this case.  \TESS\ is a much smaller instrument than JWST with smaller effective area.  Using noise values found in the ``TESS Observatory Guide'' \citep{TESSOG}, we were able to determine the expected error in two minute cadence data when observing Kepler-10, which has an $I_c$ magnitude$\sim$10.35.  In this case 52 transits would be required to detect our standard case.  Although we hope for an extended \TESS\ mission, a 34 year mission lifetime seems unlikely.   

We assume that for an equivalent planet and star, a Clarke exobelt at $10 R_p$ would produce a similar transit signature as the cloud. Based on this premise we varied the central absorptance of our simulation until the belt's central transit depth matched that of the cloud, at $\chi_0 = 0.009839$. At this value, the belt's total absorptance is the same as that for the reference cloud simulation (equal to the planet's cross-sectional area). 

\begin{figure}[tbp]
\epsscale{0.9}
\ifpdf
\plotone{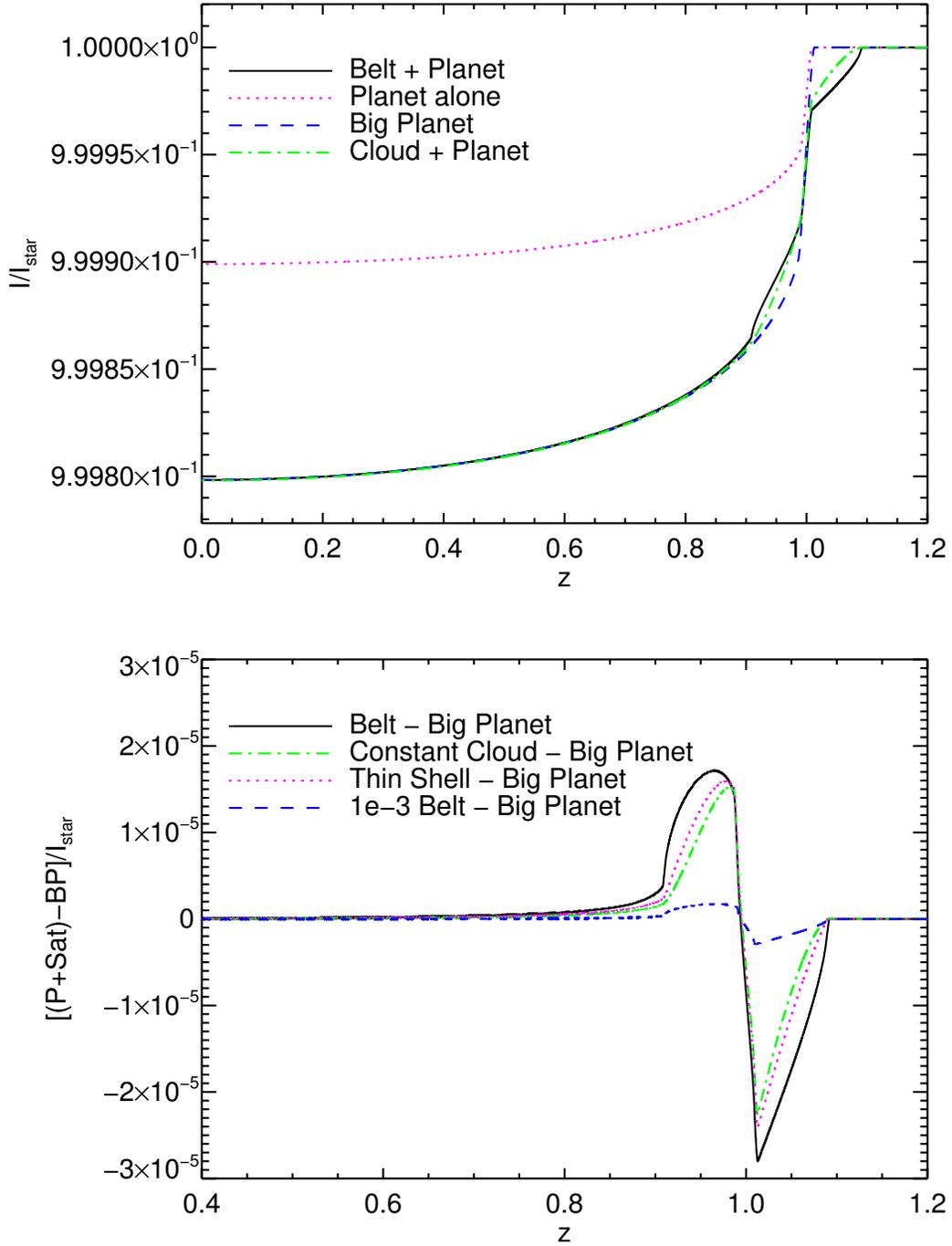}
\else
\plotone{G2_HZin_belt_col.eps}
\fi
\caption{Simulated transits for a planet at the inner edge of the HZ of a G2 star with half-solar abundance. {\it Top panel:} Transit light curves for an Earth analog with a 10$R_p$ belt with $\chi_0$ = 0.009839 (solid, G2HZin), a larger isolated planet causing a transit of the same depth (dashed), the Earth analog without a belt (dotted), and the Earth analog planet with the reference ``just detectable" constant-absorptance cloud of \KSG\ (dash-dot). {\it Bottom panel:} Shows how the G2HZin belt (solid) and cloud (dash-dot) light curves differ from the big planet light curve (relative to the stellar intensity), plus the equivalent residuals for a 10$R_p$ thin spherical shell of satellites (dotted) and a 10$R_p$ belt with absorptance $10^{-3}$ (dashed).}
\label{fig:detectable}
\end{figure}

The top panel of Figure \ref{fig:detectable} compares the light curve for this Clarke exobelt with that of the reference cloud from \KSG. The horizontal axis 
($z = d/R_{\rm star}$) is the normalized separation of the star and planet centers. Observationally, the effects of the orbiting satellites can only be detected by comparing with the light curve of an isolated planet producing a transit of the same depth (labeled `Big Planet'). These differences are shown in the bottom panel of Figure \ref{fig:detectable} for the Belt and cloud shown in the top panel. The exobelt residual signature is very similar in magnitude and shape to that of the reference cloud, so it would also be just detectable with $\sim 31$ \JWST\ transits.

For comparison, the residuals are also shown for a thin shell of orbiting mirrors with the same effective area as the constant-absorptance reference cloud. We have also calculated the transit for a belt with central absorptance $\chi_0 = 10^{-3}$ (not shown in top panel), and show the residuals in the bottom panel. This is comparable to the transit shown in Figure 4 of \SN\ (with slightly different belt inclinations and heights). It is apparent that \SN's ``observable threshold" of $\chi_0 \sim 10^{-4}$ is overly optimistic and would require thousands of transit observations to detect. Much of this discrepancy results from the simplified observing assumptions of \SN, and the fact that his simulation was for a rare G2 star only 10 light years away. 

Given the small differences in the simulations shown here, the three minimally detectable technosignature situations would be very difficult to distinguish observationally, unless their synchronous orbits make natural explanations unlikely. In addition, discriminating between rings, belts, and satellite clouds will be quite challenging based on light curve residuals alone As noted earlier, for planets tidally locked into synchronous rotation with their star, no synchronous orbits exist. So we note that even if a satellite belt could be distinguished from a spherical cloud of satellites, the expected location of a satellite belt cannot be predicted based on the planet's orbital (and presumed rotation) period, so extent cannot be used to distinguish between planetary rings and Clarke exobelts.

A ring's absorptance profile could be nearly mimicked by a thick belt or shell of satellites, but given the difficulty of detecting the overall signature, subtle differences between the various situations will be nearly impossible to detect without numerous transit observations. Multi-wavelength data would help, because satellite and ring absorptances would have different wavelength dependence. Whether such observations will be made with highly oversubscribed telescopes is likely to depend on the incidence of unusual transit light curves.

Note that a belt $10 R_p$ away from an Earth analog would be synchronously orbiting for a planet rotation period of 44.5 hours (1.86 days).  At HZin (0.75 AU) around a G2 star, the orbit period is 237 days so, like Earth, this planet is not synchronously rotating (by definition). Because the L1 point for such a situation is 176 Earth radii from the planet, satellites in a 10 $R_p$ orbit would be relatively stable.

\subsubsection{Earth's belt}\label{sect:Earthbelt}

Our second simulation highlights what would be required to make Earth's geosynchronous belt detectable. We simulate transits of Earth orbiting 1 AU from a solar-abundance G2 star, surrounded by a geosynchronous belt with $R_{belt} = 6.6R_p$. We assume the satellite density causes a central absorptance $\chi_0 = 0.02347$, so the belt produces the same transit depth as the just-detectable, constant-absorptance, $10 R_p$ reference cloud discussed in Section \ref{sect:G2HZin}. The central absorptance is larger than for our G2HZin simulation because this belt has a smaller spatial extent, so must be denser to achieve the same total absorptance (effective area).

\begin{figure}[tbp]
\epsscale{0.9}
\ifpdf
\plotone{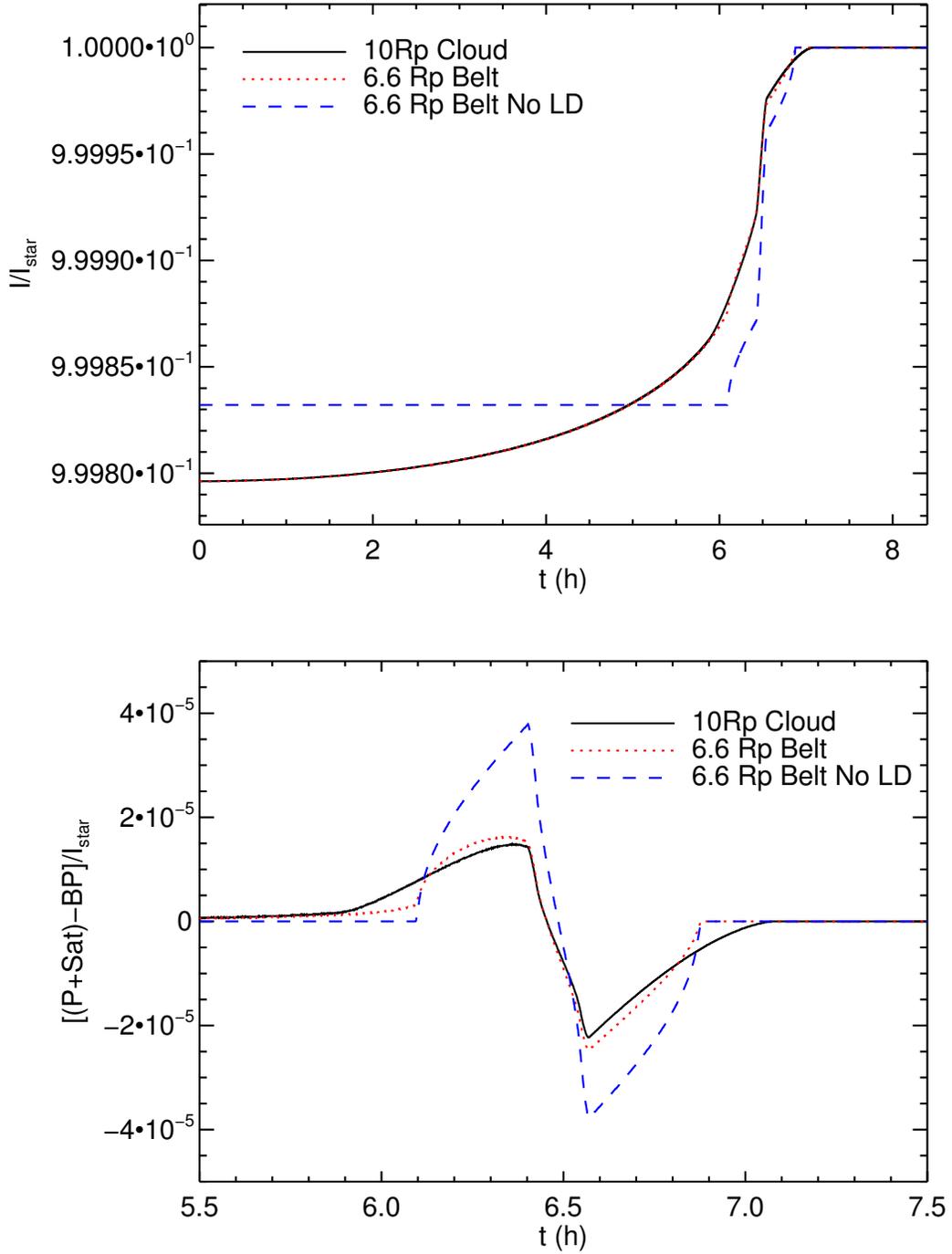}
\else
\plotone{Earth_fig_col.eps}
\fi
\caption{Simulations for Earth orbiting 1AU from a G2 solar-abundance star. {\it Top panel:} Light curves for a constant-absorptance $10 R_p$ cloud with the same properties as Figure \ref{fig:detectable} (solid), along with a 6.6 $R_p$ belt with absorptance 0.02347 chosen to match the transit depth of the cloud (dotted). The transit light curve for the same belt is also shown for a simplified star with no limb-darkening (dashed). {\it Bottom panel}: For the same situations, shows the difference near the edge of the transit between planet+belt transit and isolated big planet transit with same transit depth, relative to stellar intensity.}
\label{fig:Earth}
\end{figure}

Figure \ref{fig:Earth} compares the two transit light curves as a function of time. The lower panel focuses on the brief time near the transit edge where these light curves differ significantly from those for an isolated big planet matching their transit depth. The magnitude of observable residuals is similar for both the reference cloud and smaller belt. Although the two situations shown here have different stellar abundances, the cloud residuals change by at most $\sim 5 \times 10^{-7}$ when a solar-abundance G2 star is used.

The 10 $R_p$ constant-absorptance cloud around Earth at 1 AU from a G2 star produces a transit 1.15 times longer than if the planet were at 0.75AU. The duration of the residuals is also 1.15 times longer, thus this event could be detected using $\sim 1.15$ times {\it fewer} transits than the situation analyzed in \KSG. Although similar in shape, the two residuals caused by the $6.6 R_p$ belt are $10/6.6 R_p = 1.52$ times shorter in duration than those caused by a $10 R_p$ belt around the same planet, so they would require $\sim 1.52$ times more transits. Since the transit residuals for the $R_m = 6.6 R_p$ belt are  only slightly larger in magnitude than those for the $R_m = 10R_p$ cloud, we combine the effects of the changes in planet location and belt size to conclude that detecting Earth's geosynchronous belt with \JWST\ would require $\sim 41$ transits. 

Figure \ref{fig:Earth} also shows the effects of the same belt if limb darkening is not included.  This transit light curve shape is very similar to that of Figure 4 of \SN\ (similar but not identical situation). The transit is not as deep because at transit center both the planet and belt cover only a small region near the center of the G2 star, and without limb-darkening, the edges of the star contribute a greater fraction of total light.  The constant stellar intensity also means the belt's transit entry effects are more pronounced and the transit reaches maximum depth sooner. The bottom panel shows that without incorporating limb-darkening, the  difference signature has sharper features, and is larger in magnitude (even relative to the big planet appropriate for that shallower transit) than with limb-darkening. This is part of why \SN\ underestimated the observable threshold for $\chi_0$.

\begin{figure}[p]
\ifpdf
\plotone{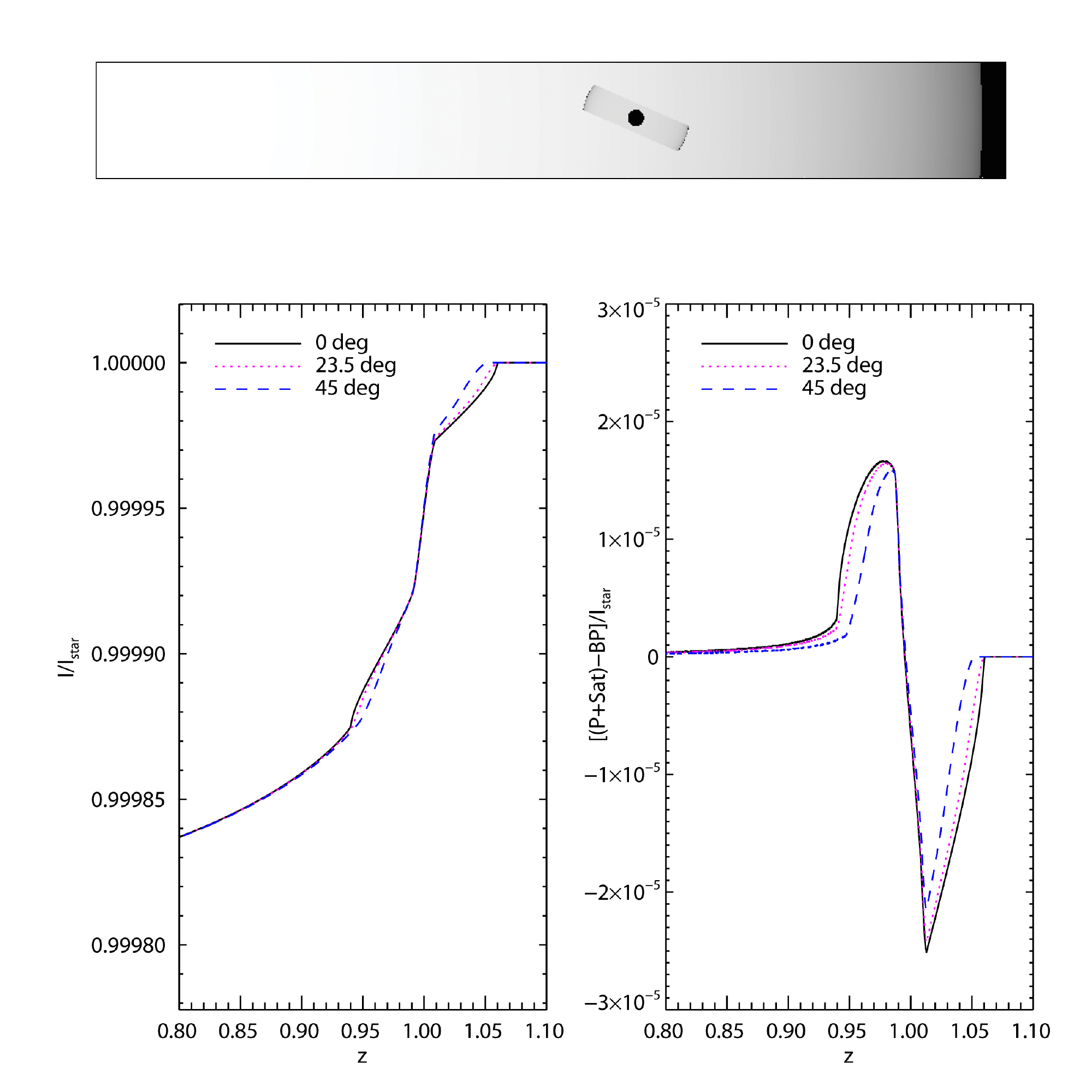}
\else
\plotone{Earth_rotate_col_new.eps}
\fi
\caption{\textit{Top panel:} Simulated image of a transit during an equinox for an Earth analog (axis tilt $23.5\degrees$) orbiting 1AU from a solar-abundance G2 star. The equatorial belt (viewed edge-on) has $\chi_0 = 0.02347$ and is located at 6.6$R_{\oplus}$. {\it Bottom Left:} Transit light curves for the same equatorial belt for three different axis tilts ($0\degrees$: solid, $23.5\degrees$: dotted, and $45\degrees$: dashed). {\it Bottom Right:} Difference of these transit light curves from that of a big planet producing a transit of the same depth, relative to the stellar intensity.}
\label{fig:rotate}
\end{figure}

Earth's spin axis is tilted relative to its orbit axis. If, from the observer's perspective, Earth transits at an equinox, the edge-on belt will be inclined from horizontal by 23.5$\degrees$, as illustrated in the top panel of Figure \ref{fig:rotate}. The bottom panels illustrate how rotating the belt in the simulation image affects the transit light curve. In general, larger axis tilts decrease the observable residuals slightly, but such effects will be very difficult to distinguish from belts with 0$\degrees$ rotation of a slightly different size and absorptance.

We also simulated belt transits which didn't pass directly in front of the stellar center. For larger impact parameters, transits are shorter in duration (less time in front of the star) and shallower (less starlight blocked due to limb-darkening). For larger impact parameters, the observable residuals (analogous to those shown in the bottom panel of Figure \ref{fig:Earth}) are significant during a greater fraction of the transit time-scale. They are also smaller in magnitude, but not substantially so until $z_{\min} \gtsim 0.75$. The trends are very similar to those seen for clouds in \KSG's Figure 7. 

\subsubsection{Practical Issues}\label{sect:practical}
How realistic is such a belt in terms of the actual satellite distribution? 
\SN\ determined the current value of $\chi_0$ for Earth's belt using a 2017 estimate for the number of satellites in geosynchronous orbit, and assuming a typical radius of 1m, a belt thickness of 150m and satellite inclinations $<15\degrees$. However, most satellites have large solar panels that will be face-on to the star during transit, so their cross-sectional areas will be larger. For example, the GOES-R satellite\footnote{From \href{http://www.goes-r.gov}{http://www.goes-r.gov}, retrieved June 06, 2018} has a cross-sectional area of $\sim 20$ m$^2$, 
which will increase the current belt to $\chi_0 \sim 2 \times 10^{-12}$. With this revision,  achieving \SN's ``detectable threshold" of $\chi_0 = 10^{-4}$ requires $3 \times 10^{10}$ satellites, at a density of $\sim 3 \times 10^{-8}$ m$^{-3}$, which corresponds to a typical separation of $\sim 310$m. Just detectable in $\sim 31$ \JWST\ transits, our G2HZin simulation with $R_{belt} = 10R_{\oplus}$ and $\chi_0 = 0.009839$ requires a satellite density of $\sim 3 \times 10^{-6}$ m$^{-3}$, or a typical separation of $\sim 67$m. 

That seems very crowded for satellites with large solar panels. The situation is even worse for our Earth-Sun analog's detectable geosynchronous belt with $\chi_0 = 0.02347$, requiring a typical separation of only 50m. This can be mitigated somewhat by utilizing a greater fraction of the geosynchronous region (defined as altitude $\pm 200$ km in the IADC Space Debris Mitigation Guidelines\footnote{From:  \href{https://www.iadc-online.org/Documents/IADC-2002-01, IADC Space Debris Guidelines, Revision 1.pdf}{https://www.iadc-online.org/Documents/IADC-2002-01, IADC Space Debris Guidelines, Revision 1.pdf}, retrieved 14 January 2019}). For an Earth-sized planet in our simulations, the 1 pixel belt thickness corresponds to 64km.  Using the larger 64km belt thickness yields typical satellite separations of $\sim 2300$m for \SN's threshold of $\chi_0 = 10^{-4}$, $510$m for our G2HZin simulation, and $380$m for a detectable belt around Earth. As a note, the total mass in satellites in these three cases (using $\sim 3000$ kg as for GOES-R), is $\sim 9 \times 10^{13}$ kg, $ 9 \times 10^{15}$ kg, and $2 \times 10^{16}$ kg.

Increasing the range of altitudes makes a greater fraction of the satellites not quite geosynchronous. Even with the revised separations, keeping these satellites from bumping into each other would require careful planning and orbit maintenance. Even for the current catalog of Earth's geosynchronous satellites and debris, the comprehensive analysis of \citet{Oltrogge18} suggests collisions {of satellites with orbiting debris may} be relatively frequent.  Their analysis implies a collision every four years for the active satellites against a catalog containing all known objects larger than 1 cm, and every 50 years against a 20cm catalog.  However, their Table 1 reveals that between 2004 and 2013, there were 9 geosynchronous satellite losses due to breakup, collision with space debris, or unknown failure, plus one likely micrometeoroid hit.  Thus the current population experiences approximately one significant event  per year, at least some of which are collisions with debris. In addition, they note that every satellite breakup results in additional debris at high relative velocities, increasing the collision likelihood. 

\citet{Friesen92} simulated the orbits of unpowered geosynchronous satellites for 100 years, and found that (on various timescales) their semimajor axes varied by less than 10 km, their eccentricities varied by $< 0.001$, but the inclinations varied by $\sim 15\degrees$. Thus inclination variations are likely to result in the largest relative velocities, and thus dominate the collision rate.
 
To estimate collision probabilities we have followed the ``gas collision" model of \citet{Oltrogge18} in the low density/low probability limit. This is merely a starting point, as ordered satellite orbits will reduce the collision rate below our estimate. For a single satellite in a Clarke belt or mean radius $r$, the collision probability in a time interval $\Delta t$ is given by
\begin{equation}
P_c=\rho V_{\rm rel} A_c \Delta t
\end{equation}
where $\rho$ is the number density of spacecraft, $V_{\rm rel}$ is the relative velocity of close encounters, and $A_c$ is the collision cross section.  For simplicity we have assumed identical spherical spacecraft with the collisions occurring whenever spacecraft centers approach within twice the spacecraft radius (i.e. $A_c=4A_s$, where $A_s$ is the spacecraft area).  In this case, $\chi_0=\rho A_s \Delta r$ where $\Delta r$ is the thickness of this belt.  

For a belt of inclination coverage $\pm \gamma$, the interaction velocity will be of order $\sin{\gamma}$ times the orbital velocity $v_o$.  Substituting, we determine that the number of collisions per unit time is
\begin{equation}
    \dot{n_c}={{4\chi_0 V_{\rm rel}}\over{\Delta r}} n
\end{equation}
where $n$ is the total number of spacecraft in the belt
\begin{equation}
n={{4\pi r^2 \chi_0  \sin{\gamma} }\over{A_s}}      
\end{equation} 
resulting in a collision rate of
\begin{equation}
\dot{n_c}={{16\pi\chi_0^2r^2 v_o}\over{A_s \Delta r}} \sin^2 \gamma
\end{equation}
This result reinforces the intuitive conclusions that higher density results in more frequent collisions, having larger spacecraft results in fewer collisions (for the same $\chi_0$), and having a greater volume in which to distribute the spacecraft reduces collision risk.

For \SN's threshold of $\chi_0 = 10^{-4}$, a geosynchronous belt thickness of 150m, and satellite inclinations $< 15\degrees$, the collision rate is $\sim 4 \times 10^8$ collisions \textbf{per second} between satellites of radius 1 meter, and $\sim 6 \times 10^7$ per second if the cross-sectional area is 20 m$^2$. The actual collision rate likely lies between these two extremes, as the cross-sectional area of solar arrays is decreased for objects approaching each other from above or below the plane.  For our minimum detectable $\chi_0 = 0.02347$ belt, the corresponding collision rate is $\sim 10^{12}$ or $10^{13}$ s$^{-1}$.  If we increase a geosynchronous belt's thickness to 64km, then collisions are less frequent by a factor of $64/0.15 = 427$.

This calculation assumes the satellites have no orbital correction capability. As a result, it predicts that Earth's 2017 population of geosynchronous satellites would collide once every 200 years or every 30 years for the smaller and larger satellites, respectively. As noted earlier, collisions occur more frequently than this, due to the presence of space debris and micrometeorites. Even for an equivalent micrometeorite and debris density,  belts with a larger number of satellites will likely experience a higher rate of significant events. A single collision in such a crowded environment would almost certainly result in a cascade of subsequent collisions.

Based on all of these considerations, disorganized belts with sufficient absorptance to be marginally detectable with current and upcoming technology are unlikely to exist without extremely sophisticated satellite steering and propulsion systems. Using our collision rates, $\dot{n_c}$, and spacecraft collision cross section, $A_c$, we can estimate the minimum delta-v necessary for a spacecraft to avoid collisions in such an environment and compare to what is currently used for geosynchronous station keeping ($\Delta v \sim 45$ m/s annually, \citet{soop94}).  In the case of the $\chi_0=10^{-4}$ belt described above, a spacecraft would require $\Delta v \sim 150$ m/s annually for collision avoidance. Spacecraft in the minimum detectable $\chi_0 = 0.02347$ belt would require $\Delta v \sim 7800$ km/s annually, which would require an extremely large amount of fuel. At constant $\chi_0$ the delta-v requirement is inversely proportional to the square of the belt thickness.  At a width of 64 km, the delta-v required is reduced to a more reasonable 45 m/s annually.  The computational burdens of tracking $\gtsim 10^{13}$ objects in orbit would still be considerable.  

We acknowledge that this is by far a worst case analysis of collision frequency, and that extrapolating from current technology is dangerous. There may exist means of mitigating the risk, for example by using different altitudes for different inclinations. An analysis of the dynamical stability of such a system would be interesting, but is beyond the scope of our current work. As satellite density increases, even such organized belts would require careful maintenance in the presence of even a small amount of other material. It is certainly possible that an advanced civilization will have means of removing micrometeoroids, dead spacecraft, and other hazards. Advanced propulsion well beyond our capabilities could potentially enable satellites to avoid collisions in a high-density environment. However if our worst-case collision estimate is accurate, no currently known technology or physics can provide the nearly unlimited $\Delta v$ required.

If an active Clarke exobelt is to be detectable, it must be carefully ordered to avoid collisions, and constantly maintained.  It is possible that debris from old space facilities could still be in orbit, even if it is no longer maintained. \citet{exobelt} suggest higher orbits could be utilized as a ``graveyard" to reduce collision probabilities with active satellites while still providing opacity to the belt. In synchronous systems it may be relatively easy to eject spent satellites into stellar orbit, although done frequently enough a significant circumstellar population may develop which could be a hazard in itself. In any case, the lifetime of this debris would likely be limited due to dynamical effects and collisions, thus would not be detectable for long relative to astrophysical time scales ($\sim$ tens of thousands of years).

\subsection{M Star Transits}

The last two situations were chosen because transits are more obvious for M stars due to their small size (although data tend to be noisier, due to their low luminosity), and because TESS is focusing on nearby stars, so will likely find exoplanets around M stars. But M8 stars are so faint that very few will be suitable for analysis with TESS \citep{wenger2000}. We chose M5 over M0 because it is the faintest class of stars for which we anticipate a reasonable number of potential targets. We further consider the relative detection requirements of various M stars in Section \ref{sect:Mdetect}.
The maximum-size belts with relatively stable satellite orbits will be the easiest to detect. Guided by the results of section \ref{sect:stability}, we chose to simulate M5 star transits for satellites orbiting $12 R_{\oplus}$ from an Earth analog at the inner edge of the habitable zone, and for satellites orbiting $33 R_{\oplus}$ from one at the outer edge of the HZ.  For both, we set $\chi_0 = 10^{-4}$.

\begin{figure}[tbp]
\epsscale{1.0}
\ifpdf
\plotone{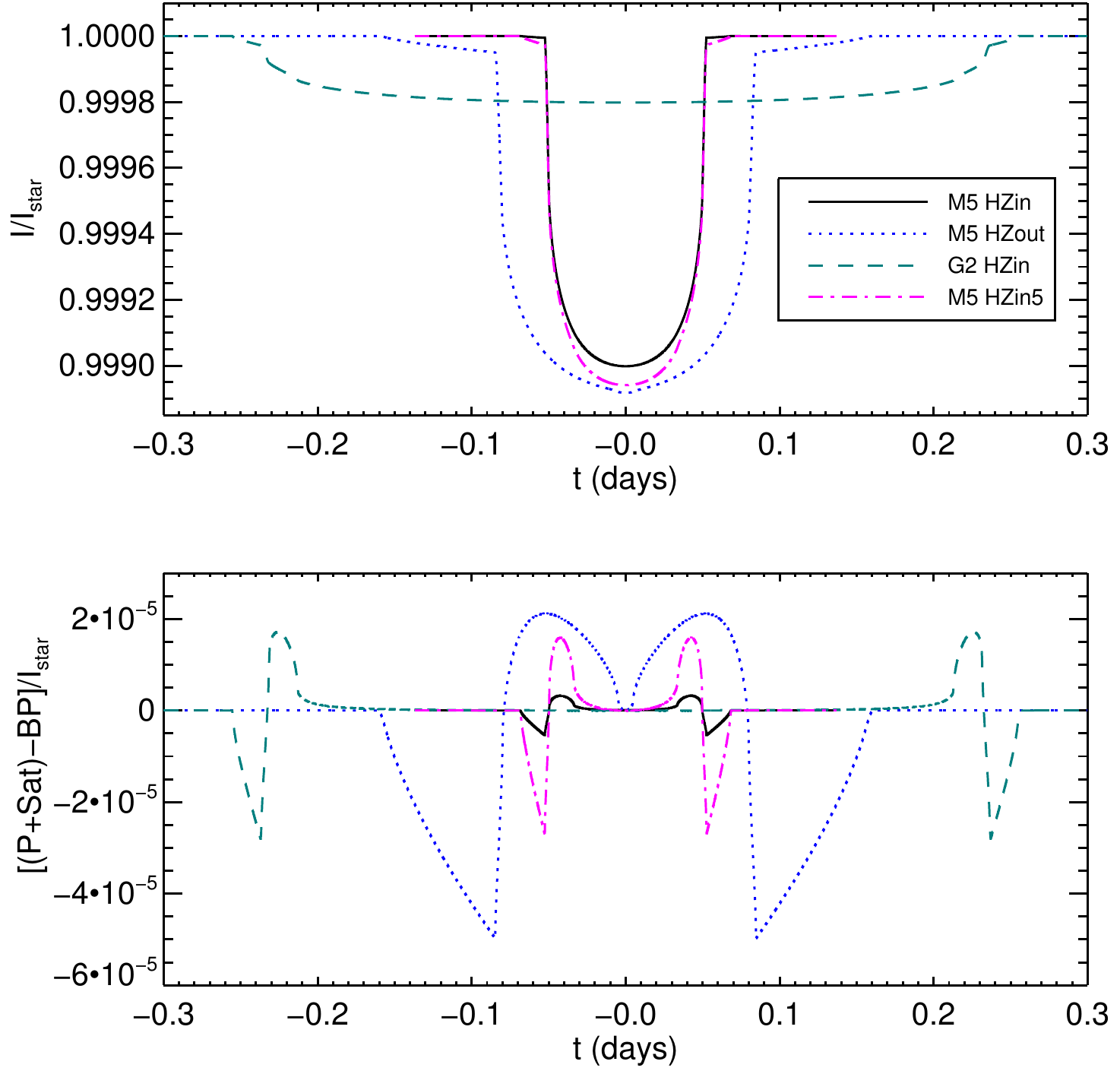}
\else
\plotone{M5_G2_vst_col.eps}
\fi
\caption{Simulated transits for several planets with exobelts, plotted against time. The top panel shows transit light curves, while the bottom panel shows how the belt light curve differs from that of a large isolated planet for each case. M5HZin (solid) simulates a belt with $\chi_0 = 10^{-4}$ and $R_{belt} = 12R_p$ around an Earth-like planet at the inner edge of the habitable zone of an M5 star. M5HZout (dotted) is the same except the belt has $R_{belt} = 33R_p$ and the planet is at the outer edge of the HZ. G2HZin (dashed) shows the same model as Figure \ref{fig:detectable}. M5HZin5 (dash-dot) is like M5HZin but with $\chi_0 = 5 \times 10^{-4}$. }
\label{fig:vst}
\end{figure}

Figure \ref{fig:vst} shows the transit light curves (top panel) and observable residuals (bottom panel) for these two situations. Both are plotted against time in days. The M5HZout transit lasts longer than the M5HZin transit, because the outer planet is moving more slowly. The M5HZout observable residuals also last a larger fraction of the transit duration, both because the planet is moving more slowly, and because the belt is larger, thus affecting the transit light curve for longer. This larger belt also results in a larger effective area (total opacity), with $A_{eff} = 0.11 A_p$ (where $A_p$ is the disk area of the planet) for M5HZout vs $A_{eff} = 0.015 A_p$ for M5HZin. Since M5HZout's larger belt has a greater total opacity for the same $\chi_0$, the resulting residuals are also much larger in magnitude. Since these largest stable belts are substantially smaller than the synchronously orbiting ones of \SN, realistic exobelts are less easily detected than suggested by that work. 

For comparison, we also plot the transit light curve for the G2HZin simulation of Section \ref{sect:G2HZin}. Its observable transit residuals are shorter in duration and smaller in magnitude than those of the M5HZout simulation. They are larger in magnitude than those for  M5HZin, and of longer duration (but only slightly). The plot also includes results for a simulation identical to M5HZin, but with a central absorptance 5 times larger, so $A_{eff} = 0.73 A_p$. The residuals of this M5HZin5 simulation are more similar to those of the G2HZin case discussed earlier, which we estimated could be just detected with 31 \JWST\ transits of data.  We further consider the detectability of exobelts for M5 transits in the next section.

We also ran M5HZin simulations assuming equinox transits of planets with identical equatorial belts, but different axis tilts. As in the Earth-Sun case described earlier, larger axis tilts decrease the magnitude and time-scale of the observable residuals somewhat, with the effects being most pronounced for the largest tilts.  Similarly, increasing the impact parameter (i.e. planet's orbit isn't quite viewed edge-on) reveals the same trends as described in Section \ref{sect:Earthbelt}.

\subsubsection{M Star Belt Detectability}\label{sect:Mdetect}

Although there are many more M than G main-sequence stars, they are much less luminous than sun-like stars. We estimated the projected count rates from G2, M0, and M5 stars using the \JWST\ Exposure Time Calculator\footnote{From \href{https://jwst.etc.stsci.edu/}{https://jwst.etc.stsci.edu}, retrieved 30 July 2019}. We used the built-in PHOENIX Stellar model \citep{Phoenix} templates for G2 (5750K), M0 (3750K), and M5 (3500K) main-sequence stars.  Each was normalized to a consistent distance using the visual absolute magnitude from \citet{zombeck}. Compared with a G2 star, the count rate is about 8 times smaller for an M0 star, and 50 times smaller for an M5 star. So for comparable detectability as our reference case, the stars would need to be at $\sim 125$ pc and $\sim 20$ pc, respectively. No built-in model was available for an M8 star, but they are even fainter, so they would have to be even closer. Thus unless more transit observations are made, exobelts such as those we've considered are un-observable except perhaps for one or two of the closest M8 stars. As noted in Section \ref{sect:G2HZin}, the situation for \TESS\ detections of such belts is even bleaker.

Because of their shorter orbital period, planets in the HZs of M stars will transit more often than those orbiting G stars, thus decreasing the overall time required to achieve the same number of transits from $\sim 20$ years for the G2HZin case to $\sim 1.3$ years for M5HZin with $\chi_0 = 5 \times 10^{-4}$. However, we must also consider that the transit duration  will differ even for identical belts, because the planet orbital speed and period do not scale the same way.  In addition, the belt's size affects the duration of its observable effects during the transit. Calculating the duration of the residuals caused by the exobelt itself allows us to estimate how the required observing time will compare with our reference case. For our M5HZin5 simulation, the belt residuals are 0.82 times shorter in duration than for G2HZin, so it will require $\sim 1.2$ times more time, or $\sim 38$ transits, assuming the exobelt absorptance causes residuals of similar magnitude.  For the M5HZout situation, belt residuals last $3.65$ times longer than G2HZin, so if they were equivalent in magnitude, it would require only $\sim$ 9 transits for detection.  The maximum stable belts for a planet at $HZ_{in}$ and $HZ_{out}$ around an M0V star are $26 R_p$ and $69 R_p$, respectively. If their absorptance produces residuals comparable in strength to our reference case, they would require 22 and 11 transits for detection.  For an M8V star, the maximum size belts of $5$ and $14 R_p$ at $HZ_{in}$ and $HZ_{out}$ would require $58$ and $27$ transits, respectively. Once again, this effect is extremely unlikely to be observable for M8 stars.

These considerations neglect the fact that detectability will also depend on the precise shape of the residuals and how sharp they are relative to the overall shape of the transit.  However we anticipate those effects will be substantially less important.

\section{Conclusions}

Detecting satellite belts in transits is very difficult.  Part of this difficulty is that the instruments involved were not designed for this task.  Detector saturation prevents observing bright sources where the signal to noise ratio would be high enough to detect the light curve variations necessary to determine the existence of satellite belts, shells or clouds in a reasonable number of transits.   If the sources are faint enough not to saturate, the SNR is lower and more transits are required.  For example, the M8.2V TRAPPIST-1 system ($I_c\sim14$) is bright enough to saturate the F150W2 band, yet faint enough to be below the \TESS\ observing limit.  Changing the band to eliminate the saturation also reduces the SNR, so is of limited value in reducing the number of transits required. Non-saturating detector technologies would help immensely, but such detectors are not currently in use.

One possible track for future work would be to define the subset of stars with known transiting planets that could be observed and to determine the number of transits necessary for various scenarios.

\acknowledgements
We thank the referee, Hector Socas-Navarro, for helpful comments and suggestions. Simulations in this paper made use of the REBOUND code which is freely available at \href{http://github.com/hannorein/rebound}{http://github.com/hannorein/rebound}. This research has made use of the SIMBAD database,operated at CDS, Strasbourg, France. During this work EJK was supported in part by donations from the Friends of SETI@home
(\href{http://setiathome.berkeley.edu}{setiathome.berkeley.edu}).  

\software{IDL Astronomy User's Library \citep{idlastro}, REBOUND \citep{ReinLiu2012}, Python 3 \citep{python, python2}, SciPy \citep{scipy},  NumPy \citep{numpy1, numpy2}, Astropy \citep{astropy},
Matplotlib \citep{2007CSE.....9...90H}, Pandas \citep{pandas}}

\end{document}

%% file: units.tex
\ifx \href \undefined \newcommand{\href}[2]{{\ttfamily #2}} \fi
\ifx \spc \undefined \newcommand{\spc}{\,} \fi
\ifpp{
\renewcommand{\doi}[1]{ DOI:\href{http://dx.doi.org/\detokenize{#1}}{\ttfamily
  \detokenize{#1}}\spc}
}
\ifx \eprint \undefined
\newcommand{\eprint}[2]{
  {#1}:\href{http://\detokenize{#1}.org/abs/\detokenize{#2}}{\ttfamily
  \detokenize{#2}}\spc}
\fi
\ifx \arxiv \undefined
\newcommand{\arxiv}[1]{ \eprint{arXiv}{#1}}
\fi
\ifx \vixra \undefined
\newcommand{\vixra}[1]{ \eprint{viXra}{#1}}
\fi
\ifx \ads \undefined
\newcommand{\ads}[1]{
  ADS:\href{http://ui.adsabs.harvard.edu/abs/\detokenize{#1}}{\ttfamily
  \detokenize{#1}}\spc}
\fi

\newcommand{\tentothe}[1]{\ifmmode {10^{#1}} \else {$10^{#1}$} \fi}
\newcommand{\tten}[1]{\ifmmode {\times 10^{#1}} \else {$\times 10^{#1}$} \fi}
\newcommand{\unit}[1]{\ifmmode {\rm\ #1} \else {$\rm #1$} \fi}

\newcommand{\TESS}{{\em TESS}}
\newcommand{\JWST}{{\it JWST}}

\newcommand{\degrees}{\ifmmode{^{\circ}}\else{$^{\circ}$}\fi}
\newcommand{\degree}{\ifmmode{^{\circ}}\else{$^{\circ}$}\fi}

\newcommand{\doublet}{\ifmmode {\lambda\lambda} \else {$\lambda\lambda$} \fi}

\newcommand{\gtsim}{\unit{_{\sim}^{>}}}

\newcommand{\lt}{\unit{<}}

\newcommand{\quarter}{\ifmmode {\frac{1}{4}} \else {$\frac{1}{4}$} \fi}

\newcommand{\singlet}{\ifmmode {\lambda} \else {$\lambda$} \fi}

\defcitealias{ksg2015}{KSG}
\newcommand{\KSG}{\citetalias{ksg2015}}
\defcitealias{exobelt}{SN}
\newcommand{\SN}{\citetalias{exobelt}}